\documentclass[prl,twocolumn,showpacs,amsmath,letter]{revtex4}
\usepackage{graphicx}
\newcommand{\Journal}[4]{#1 \textbf{#2}, #3 (#4)}

\begin{document}

\title{Dynamical Regimes Induced by Spin Transfer in Magnetic Nanopillars.}
\author{Weng Lee Lim}
\author{Andrew Higgins}
\author{Sergei Urazhdin}
\affiliation{Department of Physics, West Virginia University,
Morgantown, WV 26506}


\begin{abstract}
We demonstrate the predicted out-of-plane precession induced by spin
transfer in magnetic nanostructures with in-plane magnetic field. We
show that other magnetic excitations have a significant effect on
the stability of the out-of plane precession, making it extremely
sensitive to the orientation of the applied magnetic field. The data
are supported with micromagnetic simulations. Our results elucidate
the relation between the excitation spectrum and the specific
dynamical behaviors of nanoscale magnets.
\end{abstract}

\maketitle

\begin{figure}
\includegraphics[width=3.3in]{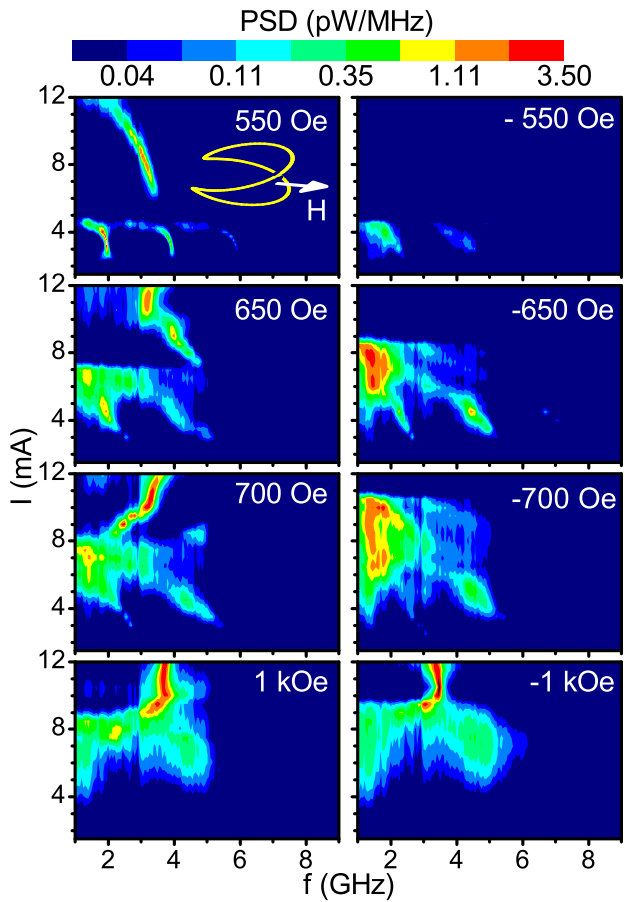}
\caption{\label{fig1} PSD vs. frequency $f$ and $I$, at the labeled
values of $H$. The logarithmic data scale brings out a break at
$2.93$~GHz of about $50$~fW/MHz due to the spectrum analyzer
crossover. Inset: clamshell trajectory of the magnetic moment, with
$H$ shown.}
\end{figure}

Spin transfer torque (ST)~\cite{slonczewski96} exerted on
nanomagnets by spin-polarized current $I$ can induce dynamical
states not accessible by any other techniques, providing a unique
opportunity to test our understanding of
nanomagnetism~\cite{cornellnature}. At $I$ just above the excitation
threshold, ST acting on the magnetic moment $m_1$ of a nanomagnet
F$_1$ causes precession on an elliptically shaped orbit, determined
by a combination of the magnetic field $H$ and the anisotropy of
F$_1$. The nanomanget is usually a thin film whose anisotropy is
dominated by the demagnetizing field, with $H$ commonly in the film
plane. The precession amplitude grows with $I$, resulting in an
orbit known as the clamshell mode (left inset in Fig.~\ref{fig1}).
Good agreement between calculations and experiments has been
achieved for precessional dynamics in this
regime~\cite{largecurrent, berkovmiltat}.

However, fast operation of magnetic devices driven by ST will likely
be performed in the less explored high-current regime. According to
simulations~\cite{berkovmiltat,lizhang,berkovgorn}, the extreme
points of clamshell eventually merge, resulting in a crossover to
the out-of-plane (OP) precession mode consisting of either the lower
or the upper half of the clamshell. Only a broad spectral feature
indicative of this mode has been seen~\cite{largecurrent},
suggesting that micromagnetic simulations may not be adequate for
the highly excited dynamical states of nanomagnets, or that
current-induced effects may not be fully described by the
established ST mechanisms.

ST also affects the magnetic layer F$_2$ used to polarize the
current, which can decrease the dynamical coherence and suppress the
OP mode~\cite{stcoupling}. Here, we report observation of the OP
precession in devices where this effect was minimized by using an
extended F$_2$. We identified and analyzed the effects of varied
direction and magnitude of $H$. The micromagnetic simulations
support our interpretation of the data, and provide insight into the
microscopic origins of the observed behaviors. Some of the results
could not be reproduced by simulations, suggesting that the
understanding of current-induced dynamics in nanomagnets is still
incomplete.

Multilayers Cu(50)Py(20)Cu(5)Py(5)Au(20), where
Py=Ni$_{80}$Fe$_{20}$ and thicknesses are in nm, were deposited on
oxidized silicon at room temperature (RT) by magnetron sputtering at
base pressure of $5\times 10^{-9}$~Torr, in $5$~mTorr of purified
Ar. F$_1$=Py(5) and about $5$~nm-thick part of F$2$=Py(20) were
patterned by Ar ion milling through an evaporated Al mask with
dimensions of $100$~nm~$\times$~$50$~nm, followed by deposition of
$30$~nm of undoped Si without breaking the vacuum. This procedure
avoids oxidation of the magnetic layers, which can affect the
magnetic dynamics~\cite{oxidation}. The mask was removed by a
combination of ion milling with Ar beam nearly parallel to the
sample surface, and etching in a weak solution of HF in water,
followed by sputtering of a $200$~nm thick Cu top contact. We
discuss data for one of three devices that exhibited similar
behaviors.

All measurements were performed at RT. The sample was contacted by
coaxial microwave probes, which were connected through a bias tee to
a current source, a lock-in amplifier, and a spectrum analyzer
through a broadband amplifier. To enable the detection of the
precessional states by electronic spectroscopy, $H$ was rotated in
the sample plane by angle $\phi=40^{\circ}$ with respect to the
nanopillar easy axis, unless specified otherwise. Positive $I$
flowed upwards. The device was characterized by magnetoresistive
(MR) measurements of its response to $H$ and $I$, yielding the
parameters essential for modeling, such as the MR of
$0.21$~$\Omega$, the dipolar coupling field of $200$~Oe caused by
the partial patterning of F$_2$, and the coercivity of $175$~Oe. The
latter was consistent with Stoner-Wohlfarth approximation,
indicating uniform magnetic reversal. The measured microwave signals
were adjusted for the frequency-dependent gain of the amplifier and
losses in the cables and probes, determined with a calibrated
microwave generator and a power meter.

Fig.~\ref{fig1} shows the dependence of the measured power spectral
density (PSD) on $H$. The spectra at $H=550$~Oe exhibit three
harmonically related peaks at $I<5$~mA, caused by the clamshell
precession. The expected fundamental frequency of the OP mode is
close to the frequency of the clamshell's second harmonic near the
crossover, since its trajectory is half that of clamshell. The
$550$~Oe data in Fig.~\ref{fig1} exhibit a spectral feature at
$I>5$~mA consistent with this relationship between the two modes.
Below, we present measurements and micromagnetic simulations that
confirm our interpretation and elucidate other, more complex
dynamical behaviors.

At $H>550$~Oe, the OP peak rapidly shifts to higher frequency.
Simultaneously, the high-current part of the peak splits from the
lower part, and eventually merges with the clamshell peaks
($H=650$~Oe and $700$~Oe data in Fig.~\ref{fig1}). One can estimate
the precession amplitude based on the total microwave power under
the peaks divided by $I^2$ (see also the description of
simulations). The largest possible emitted power for hypothetical
oscillations between the parallel (P) and antiparallel (AP)
configurations of the magnetic layers is $37.7$~pW/mA$^2$ for our
sample. The power generated by clamshell precession at $H=700$~Oe,
$I=3.5$~mA is $11.2$~pW/mA$^2$, and by the OP mode at the same $H$
and $I=10$~mA is $9.9$~pW/mA$^2$. Both values correspond to the
in-plane precession angle exceeding $90^\circ$, providing a strong
evidence for our interpretation of the spectral peaks as large-angle
precessional modes. If these spectral features were induced by
inhomogeneous dynamics rather than precession, they would result in
significantly smaller microwave power emission. The first clamshell
harmonic exhibits the smallest FWHM of $30$~MHz at $H=550$~Oe,
$I=3.5$~mA. The peaks decrease in amplitude and broaden with
increasing $H$. These behaviors suggest increasingly inhomogeneous
dynamics, resulting in decoherence of precession. In contrast, the
intensity of the OP peak at $I=10$~mA increases from $0.86$~pW/MHz
at $H=550$~Oe to $2.5$~pW/MHz at $1$~kOe, while the FWHM remains
approximately constant at $140\pm 15$~MHz.

The spectra are asymmetric with respect to the direction of $H$, as
illustrated by the difference between the left and the right panels
in Fig.~\ref{fig1}.  The clamshell peaks are consistently broader at
$H<0$ than at $H>0$. At large $I$, they are replaced by incoherent
noise rather than the OP mode. The direction of $H$ for which the OP
mode was observed varied among the samples, indicating extrinsic
origin of asymmetry. At large $H$, the data became similar for
positive and negative $H$. In particular, the $H=\pm 1$~kOe data
exhibit a sharp OP peak at $I\ge 9$~mA.

To gain insight into the origin of the behaviors shown in
Fig.~\ref{fig1}, as a well as other results discussed below, we
performed micromagnetic simulations with OOMMF open code
software~\cite{oommf}. The simulations included the current-induced
ST and Oersted field effects, but neglected thermal fluctuations.
The dipolar field was accounted for by subtracting $200$~Oe from the
in-plane component of $H$. The cell size was $4\times 4\times
5$~nm$^3$. Reducing the size to $2\times 2\times 5$~nm$^3$ did not
significantly affect the results. The parameter values available
mostly from MR measurements~\cite{bassreview} were used: current
polarization $p=0.7$, Py exchange stiffness $A=1.3\times 10^{-6}$
erg/cm, Gilbert damping $\alpha=0.03$, and the ratio $\Lambda=1.3$
of the ST magnitudes near the AP and the P states. Saturation
magnetization $M=750$~emu/cm$^3$ of Py was determined by
magnetometry of a Py(5) film prepared under the same conditions as
the nanopillar.

The spectra were calculated from the simulated time-dependent
magnetization distribution of the nanopillar. The calculated
time-dependent resistance was $R(t)=R_0+\Delta R(t)$, where
$R_0=(R_P+R_{AP})/2$, and $\Delta
R(t)=(R_P-R_{AP})\langle\mathbf{s_1\cdot s_2}\rangle/2$. Here,
${\mathbf s}_1(t), {\mathbf s}_2(t)$ are the local normalized
magnetizations of F$_1$ and F$_2$, and $\langle\rangle$ denotes
averaging over the simulation grid. The ac voltage on the input of
the amplifier $V(t)=\frac{I\Delta R(t)}{1+R_0/50\Omega}$ was
obtained by assuming that a constant current $I$ is distributed
between a $50$~$\Omega$ load and the sample resistance $R(t)$. Fast
Fourier transform (FFT) of $V(t)$ over a period of $T=16.4$~ns with
a $1$~ps step was performed after relaxation for $10$~ns. The power
spectral density was determined by $PSD(f)=2V^2(f)/(50\Omega\Delta
f)$, where $\Delta f=1/T$, and a factor of 2 accounts for the
negative-$f$ contribution to the FFT.

\begin{figure}
\includegraphics[width=3.3in]{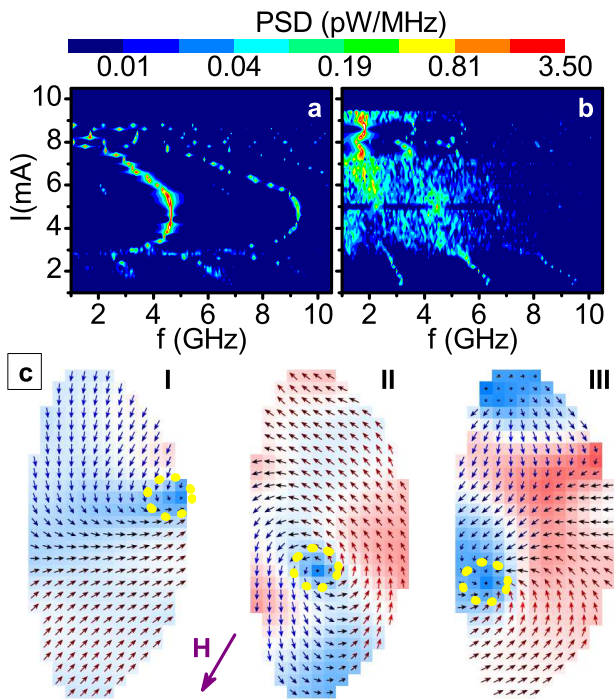}
\caption{\label{fig2} (a) Calculated PSD for $H=650$~Oe. (b) same as
(a), for $H=-650$~Oe. {\bf c}, instantaneous magnetization
distribution for $H=-650$~Oe, $I=6$~mA, captured in the I-II-III
sequence with a $100$~ps interval, showing the nucleation,
propagation, and annihilation of a vortex; the motion of its core is
marked with yellow dots. The intensity of blue(red) reflects the
out-of-plane component of the magnetization above(below) the plane.
Arrow shows the direction of $H$.}
\end{figure}

Our calculations showed that a significant asymmetry of spectra with
respect to the direction of $H$ can be induced by the simultaneous
effects of the current-induced Oersted field and a modest asymmetry
of the sample geometry. Fig.~\ref{fig2} shows results for a
nanopillar approximated by two semi-ellipses with minor semi-axes of
$34$~nm and $18$~nm, and a major axis of $104$~nm. Coherent OP
precession was obtained at $H>0$, but was suppressed at $H<0$ by
vortices and finite-wavelength spin-waves~\cite{leedieny}. The
vortices usually nucleated at the right upper edge of the
nanopillar, and annihilated at the lower left edge
(Fig.~\ref{fig2}c). The chirality of the vortices coincided with the
direction of the Oersted field's rotation, indicating that the
asymmetry of spectra is caused by the suppression or enhancement of
vortex nucleation due to the interplay of sample shape asymmetry and
the effect of Oersted field.

Simulations could not reproduce several features of the data for any
reasonable variations of nanopillar shape, distribution of current
and its polarization, or Py stiffness. Firstly, the simulated OP
peak did not exhibit the rapid shift and splitting with increasing
$H$ seen in data. Secondly, despite a significant asymmetry of the
calculated spectra, they did not reproduce the region at $I>5$~mA
where sharp OP peak was present for $H=500$~Oe, but no dynamical
features appeared at $H=-550$~Oe. The simulations also indicated
that the OP mode should exhibit multiple spectral harmonics, while
only one or two harmonics could be seen in data, regardless of the
large amplitude of precession established from the analysis of peak
intensity. These features may be caused by additional effects of
spin-polarized current neglected by the model, or by the dynamical
states not described by micromagnetic simulations.

The differences between data and simulations also open the
possibility that the high-current spectral feature in
Fig.~\ref{fig1} is caused by dynamics different from the OP mode.
Current-induced excitation of the polarizing layer F$_2$ can lead to
microwave peaks, appearing above the onset current determined by the
ratio of the volumes of F$_2$ and F$_1$~\cite{largecurrent}.
However, the effective volume of the extended layer F$_2$ in our
samples far exceeds that of F$_1$, and thus cannot explain the onset
current that at $550$~Oe is only $2.6$ times larger than the onset
of the clamshell precession. The nanopillar shape imperfections can
also result in precession around a configuration intermediate
between the AP and P states. However, this intermediate state would
quickly become unstable at increased $H$, contrary to the high-field
data in Fig.~\ref{fig1}. Such a state would also likely appear as an
intermediate-resistance step not seen in our dc measurements of MR.

\begin{figure}
\includegraphics[width=3.25in]{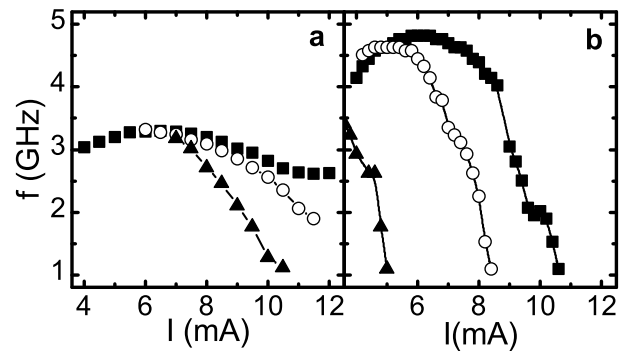}
\caption{\label{fig3} Frequency of the OP peak {\it vs} $I$ for
$\phi=48^{\circ}$ (squares), $\phi=40^{\circ}$ (open circles),
$\phi=24^{\circ}$ (triangles). (a) Measurements performed at
$H=550$~Oe, (b) Calculations performed at $H=650$~Oe, with the same
notations.}
\end{figure}

Both measurements and simulations showed that the OP precession is
extremely sensitive to the orientation of $H$, when the latter was
rotated in the film plane or tilted out of plane. Rotating $H$ in
the plane changed the dependence of the OP mode frequency on $I$,
Fig.~\ref{fig3}(a). At $\phi=48^{\circ}$, the peak exhibited a blue
shift up to $6.5$~mA, above which it gradually red shifted. At
smaller values of $\phi$, the peak broadened, decreased in
intensity, and red shifted. The correlation between the width of the
OP peak and the dependence of its frequency on $I$ is consistent
with published simulations~\cite{cornellnature,berkovgorn}. Namely,
blue shift is always predicted in the macrospin approximation, which
is more applicable to narrow coherent peaks. In contrast, the peaks
can red shift in micromagnetic simulations of the more inhomogeneous
dynamics associated with broader spectral peaks. The red shift
originates from the decrease of the total magnetic moment of the
nanopillar caused by the inhomogeneity. This interpretation was
supported by our simulations (Fig.~\ref{fig3}(b)), where decreasing
$\phi$ resulted in increasingly inhomogeneous OP dynamics. The OP
mode red shifted with $I$ at small $\phi$. At larger $\phi$, it blue
shifted at small $I$ and red shifted at large $I$, in excellent
overall agreement with the data. A somewhat larger $H$ was used in
simulations to destabilize the static current-induced AP state at
large $I$ (see discussion of Fig.~\ref{fig1}).

\begin{figure}
\includegraphics[width=3.3in]{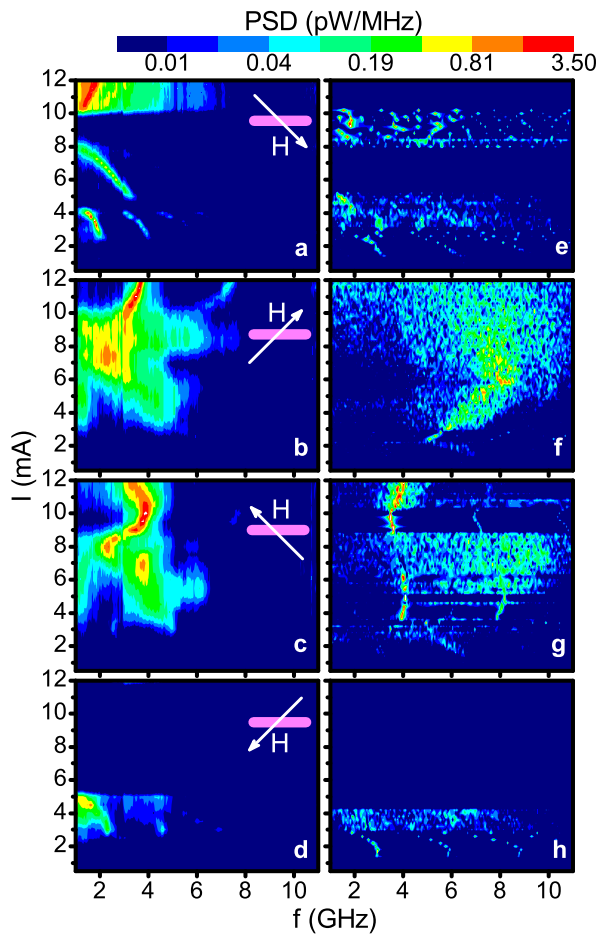}
\caption{\label{fig4} PSD for $|H|=900$~Oe tilted by $45^{\circ}$
with respect to the sample plane, at $\phi=40^{\circ}$. Insets are
side view schematics for the orientation of $H$ with respect to the
pillars, where positive in-plane direction of $H$ is to the right.
(a)-(d) Measurements, (e)-(h) Simulations.}
\end{figure}

For $H$ tilted with respect to the sample plane, the dynamics showed
dependence not only on the sign of $H$, but also on its tilting
direction (Fig.~\ref{fig4}). The component of $H$ perpendicular to
the film plane was significantly smaller than the demagnetizing
field of $9.4$~kOe, resulting in only $5^{\circ}$ tilting of $m_1$
when it was static. The OP mode appeared for both directions of $H$
above the film plane, and was suppressed for $H$ below the film
plane. The different response of dynamics to the opposite directions
of tilting indicates a significant asymmetry of this state with
respect to the film plane. This asymmetry characteristic of the OP
dynamics was reproduced by the simulations, as shown in panels
(e)-(h). Simulations for $H$ tilted above the sample plane showed
predominantly coherent OP precession {\it below the plane}, in
excellent agreement with robust spectral features seen at large $I$
in the data of Figs.~\ref{fig4}(b,c). We note that simulations with
in-plane $H$ also showed OP precession below the plane, determined
by the relative orientations of $H$ and the nanopillar easy axis. In
contrast, simulations with $H$ tilted below the sample plane showed
OP precession above the sample plane suppressed by dynamical
inhomogeneities. These results are consistent with the data.

In summary, coherent current-induced dynamics in magnetic
nanopillars was achieved by employing an extended polarizing layer.
A high-current spectral peak was identified as the out-of-plane
precession, whose dependence on the orientation of $H$ was in
excellent agreement with micromagnetic simulations. The dynamics
exhibited a significant asymmetry with respect to the direction of
$H$, attributed to a combination of nanopillar shape asymmetry and
Oersted field.

This work was supported by the NSF Grant DMR-0747609 and a Cottrell
Scholar award from the Research Corporation.

\end{document}